\documentclass[prd,superscriptaddress,showpacs,twocolumn,amsmath,amssymb]{revtex4}
\usepackage{epsfig}
\usepackage{epsf}
\usepackage{CJK}
\usepackage{amsmath}
\usepackage{tikz}
\usepackage{comment}
\usepackage{amssymb,bm,mathrsfs,bbm,amscd}
\usetikzlibrary{arrows,shapes,positioning}
\usetikzlibrary{decorations.markings}
\tikzstyle arrowstyle=[scale=1]
\tikzstyle directed=[postaction={decorate,decoration={markings,
    mark=at position .65 with {\arrow[arrowstyle]{stealth}}}}]
\tikzstyle reverse directed=[postaction={decorate,decoration={markings,
    mark=at position .65 with {\arrowreversed[arrowstyle]{stealth};}}}]

\usepackage{graphicx}

\def \jp {J/\psi}

\def \ee {e^+e^-}

\def \half {{1\over 2}}
\def \ee {e^+e^-}
\def \jp {J/\psi}

\begin{document}
\title{Polarization in $\Xi_c^0$ decays}
\author{Tong-Zhu Han}\email{hantz@ihep.ac.cn}\affiliation{ Key Laboratory of Nuclear Physics and Ion-beam Application (MOE) and Institute of Modern Physics, Fudan University, Shanghai 200443, China.}\affiliation{Physics department, Liaoning University, Shenyang 110036, People's Republic of China.}
\author{Rong-Gang Ping}\email{pingrg@ihep.ac.cn}\affiliation{ Institute of High Energy Physics, Chinese Academy of Sciences,\\
 P.O. Box 918(1), Beijing 100049, China.}\affiliation{University of Chinese Academy of Science, Beijing 100049, China.}
\author{Tao Luo}\email{luot@fudan.edu.cn}\affiliation{ Key Laboratory of Nuclear Physics and Ion-beam Application (MOE) and Institute of Modern Physics, Fudan University, Shanghai 200443, China.}
\author{Guang-Zhi Xu}\affiliation{Physics department, Liaoning University, Shenyang 110036, People's Republic of China.}

\begin{abstract}
Measurements on the decay asymmetry parameters of charmed baryons, {\it e.g.} $\Xi_c$, provide more data to test the $W$-emission and $W$-exchange mechanisms controlled by the strong and weak interactions. Taking advantage of the spin polarization in charmed baryon decays, we investigate the possibility to measure the weak decay asymmetry parameters in the $e^{+}e^{-}\to \Xi_c^0\bar\Xi_c^0$ process. We analyze the transverse polarization spontaneously produced in this process and the spin transfer in the subsequent $\Xi_c$ decays. The sensitivity to measure the asymmetry parameters are estimated for the decay $\Xi_c\to\Xi \pi$.
\end{abstract}
\maketitle

\section{Introduction}
Evidence for the charmed baryons $\Xi^+_c$ and $\Xi^0_c$ were reported for the first time by a hyperon beam experiment at CERN \cite{cern} and subsequently confirmed by other experiments \cite{Barlag:1989sw,Alam:1989qt,Albrecht:1990zk,Edwards:1995xw,Frabetti:1998kr}. Recently, precise measurements of the masses are coming from the hadron collider experiments \cite{Aaltonen:2014wfa,Aaij:2014esa}. Some exclusive decay modes are established by experiment, but its fundamental properties, such as the spin and parity, are still unconfirmed. Their quark contents are assigned as $usc$ for $\Xi^+_c$ and $dsc$ for $\Xi_c^0$ baryons. In $SU(4)$ quark model, these two states are assigned as the 20-plet with $SU(3)$ octet. Hence its spin and parity are assigned as $\half^+$.

The charmed baryon decays are suggested to be a unique laboratory to study the strong and weak interactions. Compared with the charm meson decays, the $W$-exchange processes are believed to make significant contributions due to the absence of color and helicity suppressions. This argument is confirmed by the recent measurement of the branching fraction of $\Lambda_c^+\to\Xi^0K^+$ \cite{Ablikim:2018bir}. However, it is difficult to make reliable calculations on this process since it involves nonfactorizable amplitudes. On the other hand, decay rates and asymmetry parameters for Cabbibo-favored decays, e.g. $\Xi^0_c\to\Xi^-\pi^+$ and $\Xi^+_c\to\Xi^0\pi^+$, are calculated by many groups \cite{sharma,zenczykowski,ivanov,korner,cheng,xu}.

The measurement on the decay asymmetry parameter would provide information on the nonfactorizable contributions to the decay. To date, it is measured only for the $\Xi_c^0\to\Xi^-\pi^+$ decay, namely,  $\alpha_{\Xi_c}=-0.6\pm 0.4$ \cite{Chan:2000kg}. However, the theoretical predictions came out with large uncertainty, falling in the range $(-0.99,-0.38)$ \cite{sharma,zenczykowski,ivanov,korner,cheng,xu}. Similar situation happens with the $\Xi_c^+\to\Xi^0\pi^+$ decay, and asymmetry parameter were calculated to be $\alpha_{\Xi_c}=1$ \cite{zenczykowski}, but other calculations predicted this parameter in the range $(-1,-0.27)$ \cite{sharma,ivanov,korner,cheng,xu} with same convention.

In this work, we motivate to analyze the $\Xi_c^0$ spin polarization and demonstrate how it is transferred to the decayed particles in the process $\ee\to\Xi_c^0\bar\Xi_c^0$. Taking advantage of the enhanced production cross section near the mass threshold, a data sample may be taken around $\sqrt s=5.0$ GeV in $\ee$ colliders, such as Super-Tau Charm Factory \cite{Bobrov:2015pfa}. The formulas are also applicable to the process $\ee\to\Xi_c^+\bar\Xi_c^-$.

\section{Transverse polarization  of $\Xi_c$ 	}
We suppose that the charmed baryon pairs are produced from the unpolarized beam $\ee$ collisions, namely $\ee\to\Xi_c\bar\Xi_c$. We assume that the collisions take place around the energy point $\sqrt s=2M_{\Xi_c}$, and its cross section may be enhanced close to the mass threshold of charmed baryon pair, which is similar to the enhancement of $\ee\to\Lambda_c^+\bar\Lambda_c^-$ cross section as observed in experiments \cite{bell:LLc}. The $Z$-boson contribution to this process is negligible due to the center-of-mass energy far away from the $Z$-boson mass. Hence the electromagnetic process dominates the cross section, and it conserves the spin parity.

In the production plane formed by the electron beam and the outgoing charmed baryon, the charmed baryons in the longitudinal direction in the production of this process. However, the charmed baryon may be polarized along the direction normal to the production plane as long as the process can acquire a phase angle difference between the electro- and magnetic-form factors. This kind of transverse polarization (TP) is originated from the tensor polarization of $J/\psi$ decays in $\ee$ collisions, and it has been theoretically studied for a long time \cite{Dubnickova:1992ii,Czyz:2007wi,Gakh:2005hh,Faldt:2016qee,Faldt:2017kgy,Lu:1996np,Lu:1995dk}. Recently, the transverse polarization of $\Lambda$ baryons has been observed in both continuum process and $\jp$ decays \cite{pingrg}.

The transverse polarization of the charmed baryon makes the subsequent weak decay to become a spin polarimeter, so that its decay asymmetry parameter can be measured by analysing the angular distributions of the decay products. The spin density matrix can be formulated as
\begin{equation}
\rho^{\Xi_c}={1\over 2}(P_0^{\Xi_c}I_0+{\vec P}^{\Xi_c}\cdot{\vec\sigma}),
\end{equation}
where $P_0^{\Xi_c}$ is a unpolarized cross section, $I_0$ is a $2\times2$ unit matrix, ${\vec P}^{\Xi_c}$ is a polarization vector. The spin density matrix is normalized as $P_0^{\Xi_c}=\textrm{Tr}[\rho^{\Xi_c}]$, which means that the degree of polarization is defined as $\mathcal{P}^{\Xi_c}_i=\textrm{Tr}[\rho^{\Xi_c}\sigma_i]/\textrm{Tr}[\rho^{\Xi_c}]~(i=x,y,z)$, associated with the Pauli matrices $\sigma_i$.

\begin{figure}[htbp]
\includegraphics [width=7cm]{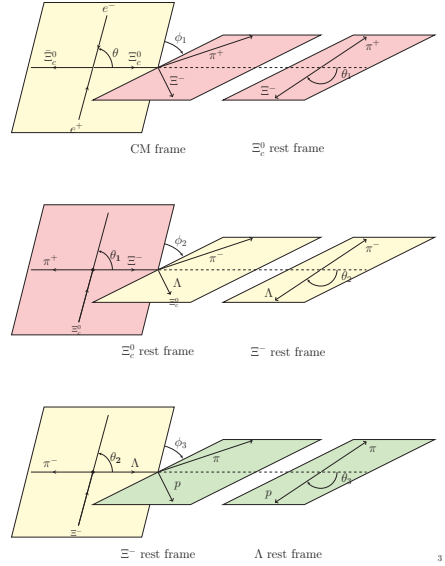}
\caption{Definition of helicity frame of $e^{+} e^{-}  \rightarrow \Xi_c^{0} \bar{ \Xi}_c ^{0},\Xi_c^{0} \rightarrow \Xi^{-}\pi^{+},\Xi^{-} \rightarrow\Lambda \pi^{-},\Lambda \rightarrow p \pi$.\label{figFrame}}
\end{figure}

Calculating the spin density matrix is straightforward. In the production plane, the orientation of the charmed baryon is denoted by a polar angle $\theta$ spanning between the positron beam and the outgoing direction of the charmed baryon, as shown in Fig. \ref{figFrame}. With helicity amplitudes defined in Table \ref{tabledef},
the elements of the $\Xi_c^0$ spin density matrix are calculated to be
\begin{eqnarray}
\rho^{\Xi_c}_{\lambda_0,\lambda_0'}&=&\sum D^{1*}_{m,\lambda_0-\lambda_1}(\phi,\theta,0)D^{1}_{m,\lambda'_0-\lambda_1}(\phi,\theta,0)\nonumber\\
&\times&A_{\lambda_0,\lambda_1}A^*_{\lambda'_0,\lambda_1},
\end{eqnarray}
where $D^{1}_{m,\lambda}(\phi,\theta,0)$ is a Wigner-$D$ function; the sum runs over the virtual photon of spin $z$-projection, $m=\pm1$. To be more specific, one has
\begin{align}
&\rho^{\Xi_c}
_{\frac{1}{2},\frac{1}{2}}=\rho^{\Xi_c}_{-\frac{1}{2},-\frac{1}{2}}=\frac{1}{2} (1+\alpha  \cos^2 {\theta }),\nonumber\\
&\rho^{\Xi_c}
_{\frac{1}{2},-\frac{1}{2}}=\rho^{\Xi_c*}
_{-\frac{1}{2},\frac{1}{2}}=-\frac{1}{4} i \sqrt{1-\alpha ^2} \sin (2 {\theta }) \sin \left(\Delta _0\right),
\end{align}
where $\alpha={|A_{1/2,-1/2}|^2-2|A_{1/2,1/2}|^2\over |A_{1/2,-1/2}|^2+2|A_{1/2,1/2}|^2}$ is the angular distribution parameter of charmed baryon, $\Delta_0$ is the phase angle difference between the two independent helicity amplitudes $A_{1/2,1/2}$ and $A_{1/2,-1/2}$. Here the spin of $\bar\Xi_c^0$ is not observed in the calculation. Then the degree of $\Xi_c^0$ polarization is calculated to be
\begin{align}\label{transpol}
\mathcal P^{\Xi_c}_{x}=&\mathcal P^{\Xi_c}_{z}=0,&\nonumber\\
\mathcal P^{\Xi_c}_{y}=& {\sqrt{1-\alpha ^2} \sin (2 {\theta }) \sin \left(\Delta _0\right)\over 2(1+\alpha  \cos^2\theta )}.
\end{align}
The results indicate that in the production $x$-$z$ plane, the polarization of the charmed baryon vanishes, but the TP component normal to the production plane emerges provided that the factor $\sqrt{1-\alpha^2}\sin(\Delta_0)$ has nonzero value.

\begin{table}[htbp]
\caption{Definition of decays, helicity angles and amplitudes, where $\lambda_i$ indicates the helicity values for the corresponding hadron.\label{tabledef}}
\begin{tabular}{lcc}
\hline\hline
Decay & Angles & Amplitude \\\hline
$e^{+} e^{-}  \rightarrow \Xi_c^{0} (\lambda_0)\bar{ \Xi}_c ^{0}(\lambda_1)$ & ($\theta,\phi$) & $A_{\lambda_0,\lambda_1}$\\\hline
$\Xi_c^{0} \rightarrow \Xi^{-}(\lambda_2)\pi^{+}$ & ($\theta_1,\phi_1$)& $B_{\lambda_2}$\\\hline
$\Xi^{-} \rightarrow \Lambda(\lambda_3)\pi^{-}$
&($\theta_2,\phi_2$)&$G_{\lambda_3}$\\\hline
$\Lambda \rightarrow p (\lambda_4)\pi$& ($\theta_3,\phi_3)$ & $F_{\lambda_4}$\\\hline\hline

\end{tabular}
\end{table}

\section{$ \Xi^{-} $ spin density matrix}
We further consider the single tag decay $\Xi_c^0\to\Xi^-(\lambda_2)\pi^+$ and $\bar\Xi_c^0$ decays to anything so that high efficiency is achived. Its helicity amplitude is denoted by $B_{\lambda_2}$, and the orientation of $\Xi^-$ is described by the helicity angles $(\theta_1,\phi_1)$. Here $\theta_1$ is defined as the angle between the $\Xi^-$ momentum in $\Xi_c^0$ rest frame and the $\Xi_c^0$ momentum in the $\ee$ center-of-mass (CMS) system, and $\phi_1$ is the angle between the $\Xi^-$ production plane and its decay plane as shown in Figure \ref{figFrame}.

The $\Xi_c^0$ weak decay gives rise to the longitudinal polarization of $\Xi^-$ by amount of $\mathcal{P}_y^{\Xi_c}\alpha_{\Xi_c}$ along the $\Xi^-$ flying direction. Hence this decay can be used to measure the $\Xi_c$ polarization degree if the decay asymmetry parameter $\alpha_{\Xi_c}$ is determined. Since the decay violates parity, any difference between the two helicity amplitudes $B_{\pm1/2}$ characterize the decay asymmetry distribution. Thus the decay asymmetry parameter is defined by $\alpha_{\Xi_c}=(|B_{+1/2}|^2-|B_{-1/2}|^2)/(|B_{+1/2}|^2+|B_{-1/2}|^2)$. This definition is consistent with the Lee-Yang parameter defined with the $S$- and $P$-wave \cite{leeyang}.

Using the obtained spin density matrix $\rho^{\Xi_c}$, the $\Xi^-$ spin density matrix is calculated by
\begin{align}
\rho^{\Xi}_{\lambda_2,\lambda_2'}=&\sum_{\lambda_0,\lambda_0'}\rho^{\Xi_c}_{\lambda_0, \lambda'_0}D_{\lambda_0,\lambda_2}^{\frac{1}{2}*}(\phi_{1},\theta_{1},0)D_{\lambda_0',\lambda_2'}^{\frac{1}{2}}(\phi_{1},\theta_{1},0)\nonumber\\
&\times B_{\lambda_2}B_{\lambda'_2}^*,
\end{align}
where $D_{\lambda_0,\lambda_2}^{\frac{1}{2}}(\phi_{1},\theta_{1},0)$ is the Wigner-$D$ function.

The further simplification yields
\begin{eqnarray}
\rho^{\Xi}_{\frac{1}{2},\frac{1}{2}}&+&\rho^{\Xi}_{-\frac{1}{2},-\frac{1}{2}}={1\over 2}(P_0^{\Xi_c}+\alpha_{\Xi_c}P_y^{\Xi_c}\sin\theta_1\sin\phi_1),\nonumber\\
\rho^{\Xi}_{\frac{1}{2},-\frac{1}{2}}&=&\rho^{\Xi*}_{-\frac{1}{2},\frac{1}{2}} =\frac{1}{4}e^{i\Delta_1}\sqrt{1-\alpha^2_{\Xi_c}}\nonumber\\
&\times&P_y^{\Xi_c}(\cos
   \theta _1 \sin \phi _1+i \cos \phi _1),
\end{eqnarray}
where $\Delta_1$ is the phase angle difference between the two helicity amplitudes $B_{1/2}$ and $B_{-1/2}$ , and the $\Xi_c$ spin polarization components are taken as
\begin{eqnarray}
P_0^{\Xi_c}&=&\rho^{\Xi_c}_{\half,\half}+\rho^{\Xi_c}_{-\half,-\half}=1+\alpha\cos^2\theta,\nonumber\\
P_y^{\Xi_c}&=&-2i\textrm{Im}(\rho^{\Xi_c}_{\half,-\half})=-{1\over 2}\sqrt{1-\alpha^2}\sin(2\theta)\sin\Delta_0.\nonumber
\end{eqnarray}

The $\Xi^-$ angular distribution is given by the trace of its spin density matrix, namely
\begin{eqnarray}\label{xixs}
\mathcal{W}_{\Xi}(\theta,\theta_1,\phi_1) &\propto &1+\alpha \cos^2\theta+\sqrt{1-\alpha ^2} \alpha _{\text{$\Xi $c}} \nonumber\\
&\times&\sin\theta \cos\theta \sin \theta _1 \sin \phi _1 \sin \Delta_0.
\end{eqnarray}
The above distribution can be understood from the role of $\Xi_c^0$ spin polarimeter, {\it i.e.}
\begin{equation}
\mathcal{W}_{\Xi}(\theta,\theta_1,\phi_1)=P_0^{\Xi_c}[ 1+\mathcal{P}_y^{\Xi_c}\alpha_{\Xi_c}\sin(\theta_1)\sin(\phi_1)],
\end{equation}
where $P_0^{\Xi_c}$ is unpolarized cross section.

The spin transfer in the $\Xi_c^0$ decay is twofold. The $\Xi^-$ transverse polarization entirely originates from the $\Xi_c^0$ transverse component, and the decayed $\Xi^-$ acquires some longitudinal polarization partly from $\Xi_c^0$ transverse polarization, and partly from the weak decay. The elements of its polarization vector are calculated to be
\small{
\begin{eqnarray}
P^{\Xi}_{x}&= &{1\over 2}\sqrt{1-\alpha_{\Xi_c}^2}P_y^{\Xi_c}(\sin\Delta_1\cos\phi_1-\cos\Delta_1\cos\theta_1\sin\phi_1),\nonumber\\
P^{\Xi}_{y}&=&{1\over 2}\sqrt{1-\alpha_{\Xi_c}^2}P_y^{\Xi_c}(\sin\Delta_1\cos\theta_1\sin\phi_1+\cos\Delta_1\cos\phi_1),\nonumber\\
P^{\Xi}_{z}&=&{1\over 2}(\alpha_{\Xi_c}P_0^{\Xi_c}-\sin\theta_1 P_y^{\Xi_c}\sin\phi_1).
\end{eqnarray}
}
\section{JOINT ANGULAR Distribution }
In order to optimally use the information available in experiment, we formulate the joint angular distributions for the full decay chain, namely, $e^{+} e^{-}  \rightarrow \Xi_c^{0} \bar{ \Xi}_c ^{0},\Xi_c^{0} \rightarrow \Xi^{-}\pi^{+},\Xi^{-} \rightarrow\Lambda \pi^{-},\Lambda \rightarrow p \pi$, and $\bar\Xi_c^0$ decaying into anything. The first two decay chains have been encoded in the $\Xi^-$ spin density matrix, and thus allow us to construct the complete decay amplitude beginning with the $\Xi^-$ decay. Helicity amplitudes for $\Xi^-$ and $\Lambda$ decays are given in Table \ref{tabledef}, and their decay asymmetry parameters are denoted by $\alpha_{\Xi}$ and $\alpha_\Lambda$, respectively, defined by
$\alpha_\Xi=(|G_{1/2}|^2-|G_{-1/2}|^2)/|G_{1/2}|^2+|G_{-1/2}|^2,$ and
$\alpha_\Lambda=(|F_{1/2}|^2-|F_{-1/2}|^2)/|F_{1/2}|^2+|F_{-1/2}|^2$.
Helicity angles, ($\theta_2,\phi_2$) for $\Xi^-\to\Lambda\pi^-$, and $(\theta_3,\phi_3)$ for $\Lambda\to p\pi^-$, are illustrated in Figure \ref{figFrame}. Then the joint angular distribution for the full decay chain is calculated by

\begin{eqnarray}
\mathcal{W}(\theta,\theta_1,\theta_2,\theta_3,\phi_1,\phi_2,\phi_3)
&\propto &\sum _{\lambda_i,\lambda_j =\pm \frac{1}{2}} \rho^{\Xi}_{\lambda_2,\lambda_2'}(\theta,\theta_1,\phi_1)\nonumber\\
&&\hspace{-3cm}\times D_{\lambda_2,\lambda_3}^{\frac{1}{2}}(\phi_2,\theta_2,0)D_{\lambda_2',\lambda_3'}^{\frac{1}{2}*}(\phi_2,\theta_2,0)\nonumber \\
&&\hspace{-3cm}\times D_{\lambda_3,\lambda_4}^{\frac{1}{2}}(\phi_3,\theta_3,0)D_{\lambda_3',\lambda_4}^{\frac{1}{2}*}(\phi_3,\theta_3,0)\nonumber \\
&&\hspace{-3cm}\times G_{\lambda_3}^*G_{\lambda_3'}|F_{\lambda_4}|^2.
\end{eqnarray}
After the $\phi_3$ angle is integrated out, then the simplified expression in terms of decay asymmetry parameters is given by
\begin{eqnarray}
\mathcal{W}(\theta,\theta_1,\theta_2,\theta_3,\phi_1,\phi_2)
&\propto &\mathcal{W}_{\Xi}(1+\alpha_\Lambda\alpha_\Xi\cos\theta_3)\nonumber\\
&&\hspace{-3.5cm}+(\alpha_\Lambda\cos\theta_3+\alpha_\Xi)(P^\Xi_x\sin\theta_2\cos\phi_2\nonumber\\
&&\hspace{-3.5cm}-P^\Xi_y\sin\theta_2\sin\phi_2+P^\Xi_z\cos\theta_2).
\end{eqnarray}

If we don't observe the angular distributions for the last step decay, we integrate out the angle $\theta_3$. Then we have a reduced distribution in terms of the $\Xi$ spin polarization as
\begin{eqnarray}
 \mathcal{W}_\Lambda(\theta,\theta_1,\theta_2,\phi_1,\phi_2)&\propto&\mathcal{W}_{\Xi}+\alpha _{\Xi}P_z^\Xi\cos\theta_2\nonumber\\
&&\hspace{-3cm}+\alpha_\Xi\sin\theta_2(P_x^\Xi\cos\phi_2-P_y^\Xi\sin\phi_2).
\end{eqnarray}

\section {Spin observable}
The transverse polarization of charmed baryon is spontaneously generated from the $\ee$
annihilation, which is characterized by Eq. (\ref{transpol}). The charmed baryon carries a reverse polarization degree in the detector of east and west region, and has a net-zero degree of polarization in the full coverage of detector. To display its distribution versus $\cos\theta$ in data analysis, a general way is to fill the distribution of $\langle \sin\theta_1\sin\phi_1\rangle$ versus $\cos\theta$, which makes it independent on the measurement of angular distribution parameter $\alpha$. Here the average is defined by
\small{
\begin{eqnarray}
\langle \sin\theta_1\sin\phi_1\rangle &=&{1\over N} \int \mathcal{W}_\Xi(\theta,\theta_1,\phi_1)\sin\theta_1\sin\phi_1d(\cos\theta_1)d\phi_1\nonumber\\ &=&{1\over 3}\mathcal{P}_y^{\Xi_c}\alpha_{\Xi_c},
\end{eqnarray}
}
with a normalization factor $$N=\int \mathcal{W}_\Xi(\theta,\theta_1,\phi_1)d(\cos\theta_1)d\phi_1.$$

\begin{figure}[htbp]
\includegraphics[width=8cm]{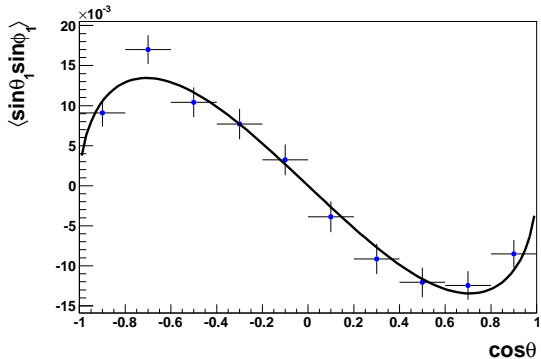}
\caption{The $\langle\sin\theta_1\sin\phi_1\rangle$ distribution versus $\cos\theta$. Dots with error bars are filled with 1.0 million MC events, and curve shows a comparison with the charmed baryon transverse polarization $\mathcal{P}^{\Xi_c}_y$.\label{toymc}}
\end{figure}

A Monte-Carlo (MC) simulation is performed to generate events using Eq. (\ref{xixs}), with
a naive choice of parameters $\alpha _{\Xi_c}= -0.1,~\alpha = 0.3$. Then we fill a histogram of $\cos\theta$ with a weight of $\sin\theta_1\sin\phi_1/N$ as shown in Figure \ref{toymc}, here $N$ is the number of generated events. With a sufficient size of MC sample, one can see that the filled distribution of $\langle\sin\theta_1\sin\phi_1\rangle$ versus $\cos\theta$ can be described with the $\mathcal{P}^{\Xi_c}_y$ distribution as given by Eq. (\ref{transpol}).

\section {Sensitivity of $\alpha_{\Xi_c}$ measurement}
To measure the decay asymmetry parameter $\alpha_{\Xi_c}$, we make use of the decay chain as much as possible. The precise measurement benefit from the full decay chain of polarization information, which is expressed in terms of hyperon $\Xi$ and $\Lambda$ weak decays. We assume that the $\alpha_{\Xi_c}$ parameter is extracted from fitting the normalized angular distribution $\mathcal{\widetilde W}(\theta,\theta_1,\theta_2,\theta_3,\phi_1,\phi_2)$, defined by $W(\theta,\theta_1,\theta_2,\theta_3,\phi_1,\phi_2)/\int...\int W(...) d\cos\theta d\cos\theta_1 d\cos\theta_2\times$ $ d\cos\theta_3d\phi_1d\phi_2$, to data event by event with a likelihood function
\begin{equation}\centering
L=\prod_{i=1}^{N} \mathcal{\widetilde W}(\theta,\theta_1,\theta_2,\theta_3,\phi_1,\phi_2,\alpha_{\Xi_c}),
\end{equation}
where $N$ is the number of observed events. The statistical sensitivity associated to the parameter estimation with the maximum likelihood method is determined by the relative uncertainty
\begin{equation}
\delta(\alpha_{\Xi_c})=\frac{\sqrt{V(\alpha_{\Xi_c})}}{|\alpha_{\Xi_c}|},
\end{equation}
where $ V(\alpha_{\Xi_c})$ denotes the variance of the parameter $\alpha _{\Xi_c }$, which can be determined by
\begin{eqnarray}
V^{-1}(\alpha_{\Xi_c})&=&N\int \frac{1}{\mathcal{\widetilde W}(\theta_{i},\phi_i,\alpha _{\Xi_c })} \left[\frac{\partial\mathcal{\widetilde W} (\theta_{i},\phi_i,\alpha _{\Xi_c })}{\partial \alpha _{\Xi_c }}\right]^2\nonumber\\
&\times& \prod_{i} {d\cos\theta_i}\prod_{j}d\phi_j.
\end{eqnarray}
To get the dependence of sensitivity on the signal yields $N$, we estimate the value of $\delta(\alpha_{\Xi_c})$ by taking the parameters \cite{pdg} as $\alpha _{\Xi }=-0.392\pm0.008, \alpha _{\Lambda}= 0.750\pm0.010,$  but no measurement is available for the $\Xi_c$ angular distribution parameter, we naively take $\alpha = 0.6$. The parameter $\alpha _{\Xi_c }$ sensitivity is calculated by follows
\begin{equation}
\frac{0.251^{+0.002}_{-0.008}}{\sqrt N} < \delta(\alpha _{\Xi_c })<\frac{1.943^{+0.066}_{-0.012}}{\sqrt N},
\end{equation}
where the uncertainties due to the $\alpha _{\Xi }$ and $\alpha _{\Lambda}$ uncertainties, and the lower and upper bonds are determined by taking the $\alpha_{\Xi_c}=-0.1$ and $-0.9$, respectively. Here uncertainties are due to the $\alpha _{\Xi }$ and $\alpha _{\Lambda}$ uncertainties.

\begin{figure}[h]\centering
\includegraphics [width=8cm]{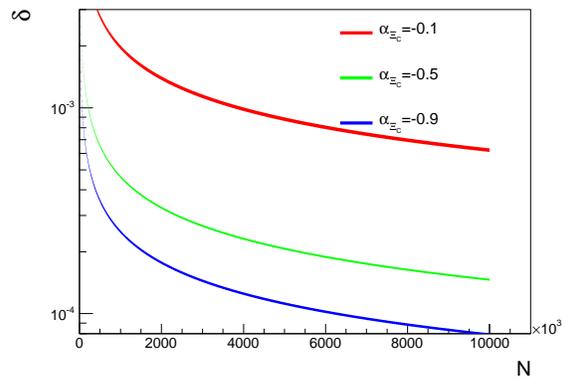}
\caption{The $\alpha_{\Xi_c}$ sensitivity versus the signal yields $N$ in terms of different value $\alpha_{\Xi_c}$. The curves from top to bottom corresponds to the $\alpha_{\Xi_c}$ value taken as $-0.1$, $-0.5$, and $-0.9$, respectively.  \label{sensitivity}
}
\end{figure}

As for other $\alpha_{\Xi_c}$ values, we plot the sensitivity versus the $\Xi_c$ statistics as shown in Fig. \ref{sensitivity}, where the experimental effects, such as the backgrounds and the detection angular acceptance, are not taken into consideration. One can see that if we take $\alpha_{\Xi_c}=-0.6\pm0.4$ for the $\Xi_c^0\to\Xi^-\pi^+$ decay, the signal 50,000 events yields sensitivity of this parameter reaching to precision of $0.1\sim0.9\%$. The $\alpha_{\Xi_c}$ measurement can be performed in the super KEKB, or in the future Super Tau Charm Facility, which is proposed by the Chinese and Russian Physicists, with the center-of-mass energy from 2 to 5 GeV.
\newpage

\section{summary}
We formulate the polarization in the process $\ee\to\Xi_c^0\bar\Xi_c^0$ motivated by study the weak asymmetry decay parameters for the $\Xi_c\to\Xi\pi$ decay. As in other continuum process, the transverse polarization may be spontaneously generated accompanied by the baryon pair production from the $\ee$ annihilations. We formulate the decay asymmetry parameter for the weak decay $\Xi_c\to\Xi\pi$, and we also show how the transverse polarization is transferred from the charmed baryon to the decayed $\Xi$ hyperon. Taking advantage of full decay chain, we formulate the joint angular distributions for the decay chain $\Xi\to\Lambda\pi$, and $\Lambda\to p\pi^-$. A Monte-Carlo simulation is performed to show how to display the transverse polarization effects. The sensitivity to measure the $\alpha_{\Xi_c}$ asymmetry parameters is estimated with the full decay chain formula.

\vspace{1cm} {\bf Acknowledgements:}  The work is partly supported by
the National Natural Science Foundation of China under Grants No. 11875262, 11805037, 11705078; Joint Large-Scale Scientific Facility Funds of the NSFC and CAS under Contracts No. U1832121, and also by Shanghai Pujiang Program under Grant No. 18PJ1401000 and Open Research Program of Large Research Infrastructures (2017), Chinese Academy of Sciences.

\end{document}